%% file: main.tex
\documentclass{aa}  

\usepackage{graphicx}
\usepackage{txfonts}
\usepackage{comment}
\usepackage[dvipsnames]{xcolor}
\usepackage{soul}
\pdfoutput=1
\definecolor{mypink}{RGB}{219,48,122}
\usepackage[tableposition=top]{caption}

\definecolor{purple}{RGB}{156,81,182}

\begin{document} 

\title{ACT-DR5 Sunyaev-Zel’dovich Clusters: weak lensing mass calibration with KiDS}

\author{Naomi Clare Robertson\inst{1,2,3}\fnmsep\thanks{naomi.robertson@ed.ac.uk}, Crist\'{o}bal Sif\'{o}n\inst{4}, Marika Asgari\inst{5,6}, Nicholas Battaglia\inst{7}, Maciej Bilicki\inst{8}, J. Richard Bond\inst{9}, Mark J. Devlin\inst{10}, Jo Dunkley\inst{11,12}, Benjamin Giblin\inst{13,1}, Catherine Heymans\inst{1,14}, Hendrik Hildebrandt\inst{14}, Matt Hilton\inst{15,16}, Henk Hoekstra\inst{17}, John P. Hughes\inst{18}, Konrad Kuijken\inst{17}, Thibaut Louis\inst{19}, Maya Mallaby-Kay\inst{20}, Lyman Page\inst{11}, Bruce	Partridge\inst{21}, Mario Radovich\inst{22}, Peter Schneider\inst{23}, HuanYuan Shan\inst{24,25}, David N. Spergel \inst{26}, Tilman Tr\"{o}ster\inst{27}, Edward J. Wollack\inst{28}, Cristian~Vargas\inst{29}, Angus H. Wright\inst{14}
}
\institute{
Institute for Astronomy, University of Edinburgh, Royal Observatory, Blackford Hill, Edinburgh, EH9 3HJ, UK
\and
Institute of Astronomy, University of Cambridge, Madingley Road, Cambridge, CB3 0HA 
\and
Kavli Institute for Cosmology Cambridge, Madingley Road, Cambridge, CB3 0HA
\and
Instituto de Física, Pontificia Universidad Católica de Valparaíso, Casilla 4059, Valparaíso, Chile
\and
E.A Milne Centre, University of Hull, Cottingham Road, Hull, HU6 7RX, United Kingdom
\and
Centre of Excellence for Data Science, AI, and Modelling (DAIM), University of Hull, Cottingham Road, Kingston-upon-Hull, HU6 7RX, United Kingdom
\and
Department of Astronomy, Cornell University, Ithaca, NY 14853, USA
\and
Center for Theoretical Physics, Polish Academy of Sciences, al. Lotników 32/46, 02-668 Warsaw, Poland
\and
Canadian Institute for Theoretical Astrophysics, University of Toronto, 60 St. George St., Toronto, ON, M5S 3H8, Canada
\and
Department of Physics and Astronomy, University of Pennsylvania, 209 South 33rd Street, Philadelphia, PA, USA 19104
\and
Joseph Henry Laboratories of Physics, Jadwin Hall, Princeton University, Princeton, NJ, USA 08544
\and
Department of Astrophysical Sciences, Peyton Hall, Princeton University, Princeton, NJ, USA 08544
\and
Instituto de Ciencias del Cosmos (ICC), Universidad de Barcelona, Martí i Franquès, 1, 08028 Barcelona, Spain
\and
Ruhr University Bochum, Faculty of Physics and Astronomy, Astronomical Institute (AIRUB), German Centre for Cosmological Lensing, 44780 Bochum, Germany
\and
Wits Centre for Astrophysics, School of Physics, University of the Witwatersrand, 2050, Johannesburg, South Africa 
\and
Astrophysics Research Centre and School of Mathematics, Statistics and Computer Science, University of KwaZulu-Natal, Durban
\and
Leiden Observatory, Leiden University, PO Box 9513, NL-2300 RA Leiden, The Netherlands
\and
Department of Physics and Astronomy, Rutgers University, 136 Frelinghuysen Road, Piscataway, NJ 08854, USA 
\and
Université Paris-Saclay, CNRS/IN2P3, IJCLab, 91405 Orsay, France
\and
Department of Astronomy and Astrophysics, University of Chicago, 5640 S Ellis Ave, Chicago, IL 60637, USA 
\and
Department of Physics and Astronomy, Haverford College, 370 Lancaster Avenue, Haverford, PA 19041, USA
\and
INAF - Osservatorio Astronomico di Padova, via dell'Osservatorio 5, 35122 Padova, Italy
\and
Argelander-Institut f. Astronomie, Univ. Bonn, Auf dem Huegel 71, D-53121 Bonn, Germany
\and
Shanghai Astronomical Observatory (SHAO), Nandan Road 80, Shanghai 200030, China
\and
University of Chinese Academy of Sciences, Beijing 100049, China
\and
Center for Computational Astrophysics, Flatiron Institute, 162 5th Avenue, New York, NY 10010, USA
\and
Institute for Particle Physics and Astrophysics, ETH Zürich, Wolfgang-Pauli-Strasse 27, 8093 Zürich, Switzerland
\and
NASA Goddard Space Flight Center, 8800 Greenbelt Rd, Greenbelt, MD 20771, USA 
\and
Instituto de Astrof\'isica and Centro de Astro-Ingenier\'ia, Facultad de F\`isica, Pontificia Universidad Cat\'olica de Chile, Av. Vicu\~na Mackenna 4860, 7820436 Macul, Santiago, Chile
}

\date{Received XX YY 20ZZ; accepted XX YY 20XX}

\abstract{We present weak gravitational lensing measurements of a sample of 157 clusters within the Kilo Degree Survey (KiDS), detected with a $>5\sigma$ thermal Sunyaev-Zel’dovich (SZ) signal by the Atacama Cosmology Telescope (ACT). Using a halo-model approach we constrain the average total cluster mass, $M_{\rm WL}$, accounting for the ACT cluster selection function of the full sample. We find that the SZ cluster mass estimate $M_{\rm SZ}$, which was calibrated using X-ray observations, is biased with $M_{\rm SZ}/M_{\rm WL} = (1-b_{\rm SZ}) = 0.65\pm 0.05$. Separating the sample into six mass bins, we find no evidence of a strong mass-dependency for the mass bias, $(1-b_{\rm SZ})$. Adopting this ACT-KiDS SZ mass-calibration would bring the {\it Planck} SZ cluster count into agreement with the counts expected from the {\it Planck} cosmic microwave background $\Lambda$CDM cosmological model, although it should be noted that the cluster sample considered in this work has a lower average mass $M_{\rm SZ, uncor} = 3.64 \times 10^{14} M_{\odot}$ compared to the {\it Planck} cluster sample which has an average mass in the range $M_{\rm SZ, uncor} = (5.5-8.5) \times 10^{14} M_{\odot}$, depending on the sub-sample used.}


\titlerunning{ACT-DR5: Cluster Mass Calibration with KiDS}
\authorrunning{Robertson \& the KiDS and ACT Collaborations et al.}
\maketitle

\section{Introduction}
\label{introduction}
The locations of galaxy clusters trace the position of peaks in the large scale matter distribution. As a result, their properties -- including number density, masses, baryon content and evolution -- contain information about the growth of structure in our Universe and can constrain quantities such as the amplitude of matter fluctuations, $\sigma_8$, the matter density, $\Omega_{\rm m}$, the sum of neutrino masses and the dark energy equation of state.
The key statistic is their number density as a function of mass and redshift, $N(M,z)$. In practice, however, we cannot measure the mass directly and instead infer it from some observable quantity that correlates with mass \citep[see for example,][for a review]{Allen2011}.

Clusters can be observed across a range of wavelengths, resulting in a variety of mass proxies. For example, they emit X-rays from the hot intra-cluster gas, with the gas temperature scaling with mass \citep[for example,][]{Ebeling1998,Bohringer2004,Vikhlinin2009}.
In the optical and infrared, the cluster richness and stellar luminosity of the galaxies within the cluster also correlate with mass \citep[for example,][]{Gladders2005,Rykoff2016}. At millimeter wavelengths clusters are detected through the thermal Sunyaev-Zel’dovich effect \citep[SZ,][]{Zeldovich1969,Sunyaev1972,Birkinshaw1984}, which occurs when CMB photons are inverse Compton scattered off electrons in the hot cluster gas. In this case, the mass observable is the volume integrated intra-cluster medium (ICM) pressure, measured with the Compton--$y$ parameter. Because the SZ effect is only weakly dependent on redshift, catalogues produced with this effect have a mass threshold that is almost constant with redshift, which makes them excellent tracers of the underlying redshift distribution.

For all cluster detection methods, using clusters for cosmology is currently limited by the uncertainty in the mass-observable scaling relations which need to be calibrated.
In the case of the SZ effect, a mass $M_\mathrm{SZ}$\footnote{
Consistent with all works referenced in this discussion, all masses refer to $M_{\rm 500c}$, i.e., the mass enclosed within a radius $r_{\rm 500c}$ enclosing 500 times the critical density of the Universe at the relevant redshift.} is usually derived from the Compton--$y$ parameter by assuming the gas and density profiles estimated by \cite{Arnaud2010}, from X-ray observations of local clusters, are in hydrostatic equilibrium \citep[e.g.,][]{Hilton2020, Hasselfield2013, Planck2016}. 

Weak gravitational lensing provides a tool to calibrate biases in the SZ mass estimate, as it is sensitive to the entire mass of the cluster, both dark and baryonic, and does not depend on its dynamical state \citep{Hoekstra2013}. Previous estimates of weak-lensing masses have been made for SZ-selected clusters from the South Pole Telescope \citep[SPT][]{mcinnes2009,high2012,schrabback2018,Dietrich2019,Stern2019}, the Atacama Cosmology Telescope \citep[ACT][]{miyatake2013,jee2014,battaglia2016,Miyatake2019} and the {\it Planck} satellite \citep{vonderlinden2014b,hoekstra2015,penna-lima2017,Sereno2017,medezinski2018, Herbonnet2020}.
Many of these studies have therefore focused on measuring the mass bias, which quantifies the cluster mass fraction left unaccounted for in the $y-M$ scaling relation. The mass bias is defined as
\begin{equation}
    (1-b_{\rm SZ}) = \frac{M_{\mathrm{SZ}}}{M_{\mathrm{WL}}} \, ,
\end{equation}
where $M_{\mathrm{SZ}}$ is the mass derived from the SZ Compton--$y$ observable, assuming the X-ray derived $y-M$ relation from \citet{Arnaud2010}, and $M_{\mathrm{WL}}$ is the mass measured using weak gravitational lensing which we take to be the `true' mass. This definition of $M_{\rm SZ}$ is not an `absolute' definition of the SZ mass, but is the one chosen for this analysis. There have been investigations into whether $(1-b_{\rm SZ})$ may depend on mass or redshift \citep{Henson2017,Remazeilles2019,Herbonnet2020}; current data are not yet conclusive given the uncertainties.
Recent estimates of this bias from SZ clusters detected using \textit{Planck} and ACT are reported in \cite{medezinski2018} and \cite{Miyatake2019}, with alternative parameterisations of the mass scaling reported from SPT in \cite{Bocquet2019}. 

In the {\it Planck} SZ cluster counts analysis \citep{Planck2016} and the {\it Planck} primary CMB analysis \citep{Planck2018} inconsistent results were obtained if hydro-static equilibrium was assumed. This difference in results could be resolved by relaxing this assumption, yielding a mass bias of $0.62 \pm 0.03$, obtained by combining the primary CMB spectra and cluster counts likelihood. 

In the analysis presented here, we estimate the mass bias for the cluster sample from ACT using weak lensing data from the Kilo Degree Survey \citep[KiDS,][]{Kuijken2019,Giblin2020}. The ACT experiment has been used to construct a sample of more than 4000 confirmed clusters over a 13\,000 deg$^2$ region, measured during the 2008-18 observing seasons \citep{Hilton2020}. KiDS provides 774 deg$^2$ of weak lensing data in the area covered by ACT; we find 157 SZ-confirmed clusters in this region which are detected with a signal-to-noise greater than 5.
We estimate the stacked masses of these clusters and the bias parameter. This allows us to probe the lower end of the SZ cluster mass range $(1.72 - 10.4) \times 10^{14} {\rm M}_{\odot}$ with a larger sample size than was previously available. This study also provides a means to cross-check consistency of data from different weak lensing surveys which have survey areas in common with ACT.

The outline of this paper is as follows. We summarise the data and observable quantities in Sect. \ref{sec:data}. Our results are presented in Sects. \ref{sec:result_stack} and \ref{sec:b} with our conclusions following in Sect. \ref{sec:conclude}. We assume a flat $\Lambda$CDM cosmology with $\Omega_{\rm m}=0.3$, $\Omega_{\Lambda}=0.7$ and $H_0=70 \, {\rm km \, s^{-1} \, Mpc^{-1}}$ throughout, unless otherwise stated. We investigate the dependence of our result on fixing cosmology when inferring the mass, which is discussed in Sect. \ref{sec:result_stack}.

\section{Data and observables}
\label{sec:data}
\subsection{The Atacama Cosmology Telescope SZ-selected Cluster Sample}
\begin{figure*}
	\includegraphics[width=\textwidth]{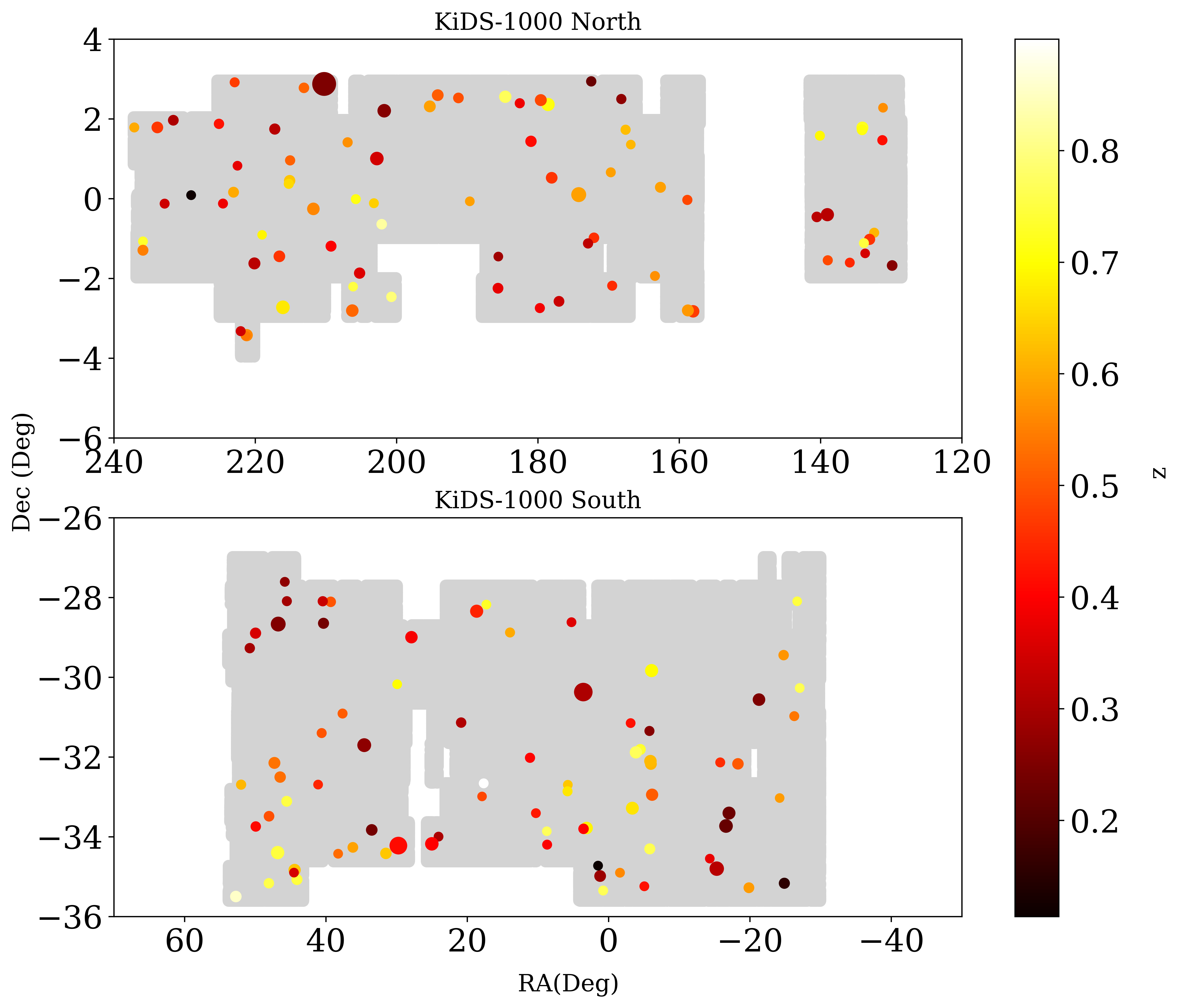}
    \caption{The KiDS North and South survey areas are shown in light grey in the upper and lower panel respectively. The positions of ACT clusters used in this analysis are indicated by each circle. The colour of the circles corresponds to cluster redshift, with yellow indicating high redshift and black indicating low redshift. The size of each circle scales with the SZ signal-to-noise, which is correlated with the cluster mass when the noise is the same everywhere, such that larger circles represent clusters with a greater signal-to-noise than smaller circles.}
    \label{fig:cluster_plot}
\end{figure*}

\begin{figure}
	\includegraphics[width=1.0\columnwidth]{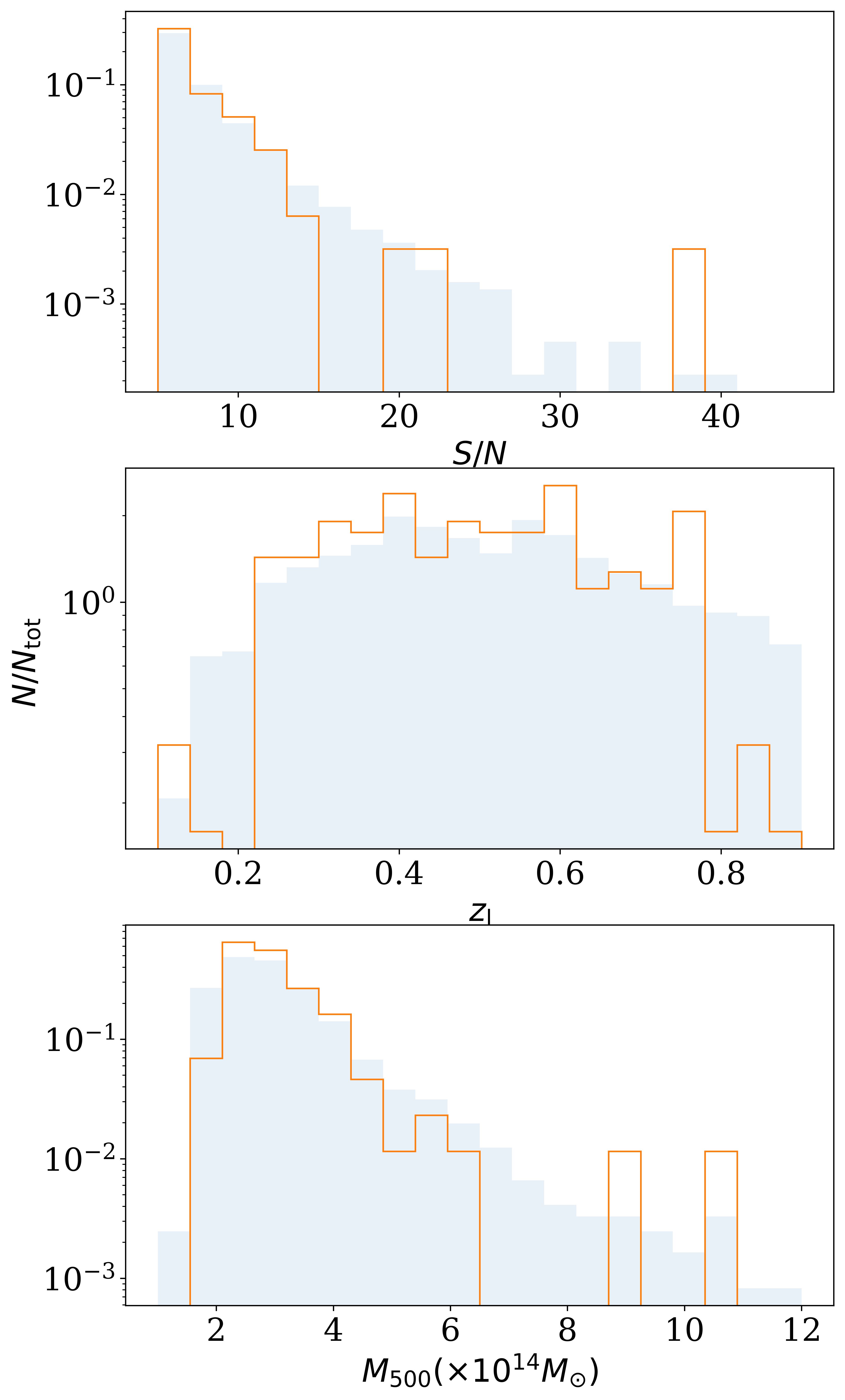}
    \caption{Upper: The normalised SZ signal-to-noise distribution of the ACT clusters found in the KiDS-1000 region. Middle: The normalised redshift distribution of these clusters. Lower: The normalised SZ-derived mass distribution of these clusters, estimated from the SZ observable Compton--$y$ values. The blue shaded region corresponds to clusters selected to have an SZ signal-to-noise greater than 5 and in the redshift range 0.1 to 0.9 across the full ACT footprint. The orange line corresponds to the selection of 157 clusters used in this analysis that lie within the KiDS footprint. 
    }
    \label{fig:redshift}
\end{figure}

We use the ACT cluster catalogue described in \citet{Hilton2020}, based on the ACT Data Release 5 (DR5) 2008-2018 coadded maps described in \cite{Naess2020}.
The cluster catalogue is constructed from coadded maps at 90 and 150 GHz \citep{Naess2020}. Clusters are identified using a suite of multi-frequency matched filters, and confirmed using imaging from the public large imaging and spectroscopic surveys and spectroscopic measurements reported in the literature \citep[see][for details]{Hilton2020}.
From this optical data, the cluster redshifts are also determined and a cluster may therefore have either a spectroscopic or photometric redshift. 
The signal-to-noise of the cluster signal is determined from a single fixed filter scale which simplifies the survey selection function. The fraction of false positives is 0.03 for clusters that are detected with a signal-to-noise greater than 5 \citep[shown in Fig. 6][]{Hilton2020} and rises to 0.34 for a signal-to-noise cut of 4. The cluster locations are shown in Fig. \ref{fig:cluster_plot}, and the distribution of their SZ detection signal-to-noise, redshifts and SZ estimated masses, are shown in Fig. \ref{fig:redshift} in comparison to the full ACT sample. Fig. 2 shows that the clusters found within the KiDS survey region are a good representation of the full ACT cluster sample \citep[see Fig. 18][]{Hilton2020}.

\citet{Hilton2020} estimate the Compton--$y$ parameter and associated SZ mass, $M_{\rm SZ}$, for each cluster; we summarise their method here. The Compton--$y$ signal is modelled using the universal pressure profile \citep{Arnaud2010} which describes the intra-cluster gas pressure, $p$, as a function of scaled radius, $x = r/r_{\rm 500c}$ where $r_{\rm 500c}$ is the radius within which the density is 500 times the critical density $\rho_{\rm c}$, at $z=0$. This relation was modelled with a generalised Navarro-Frenk-White (NFW) profile from \citet{Nagai2006} as
\begin{equation}
    p(x) = \frac{p_0}{(c_{500}x)^{\gamma}[1+(c_{500}x)^{\alpha}]^{(\beta-\gamma)/\alpha}} \, ,
\end{equation}
where $c_{500}$ is the concentration parameter and $\{\gamma,\alpha,\beta\}$ describe the slope at different radii. This includes any mass dependence in the profile shape and has been calibrated to X-ray observations of nearby clusters. From these assumptions, following \cite{Arnaud2010}, the cluster central Compton parameter, $y$, can be related to the SZ mass via 
\begin{equation}
    \tilde{y}_0 = 10^{A_0} E^2(z) \, \left( \frac{M_{\rm SZ,500}}{M_{\mathrm{piv}}}\right)^{1+B_0} Q(M_{\rm SZ,500},z) \, f_{\mathrm{rel}}(M_{\rm SZ,500},z) \, , 
\end{equation}
where $10^{A_0}=4.95\times 10^{-5}$ is the normalisation, $E(z)$ is the ratio of the Hubble constant at a redshift $z$ to the present value, $B_0=0.08$ is a scaling parameter and $M_{\mathrm{piv}}=3 \times 10^{14} \mathrm{M}_{\odot}$ is a mass normalisation parameter. The cluster-filter scale mismatch function $Q$ accounts for the discrepancy between the size of a cluster with a different mass and redshift to the reference model used to define the matched filter, which includes the effects of the beam \citep[see][for full details]{Hasselfield2013}.

The signal we observe, at a projected angle $\theta$ from the cluster centre, due to the SZ effect is a change in radiation intensity which can be expressed in terms of CMB temperature by
\begin{equation}
    \frac{\Delta T(\theta)}{T_{\mathrm{CMB}}} = f_{\mathrm{SZ}} \, y(\theta) \, .
\end{equation}
In the non-relativistic limit, $f_{\mathrm{SZ}}$ only depends on the observed radiation frequency. In \citet{Hilton2020} a correction factor $f_{\mathrm{rel}}$ is applied to include relativistic effects for gas temperature, as determined in \cite{Itoh1998}.

To obtain the cluster mass, the intrinsic scatter $\sigma_c$ in the $y-M$ scaling relation must be accounted for. \cite{Hilton2020} assume a log-normal distribution with dispersion $\sigma_c=0.2$. This level of scatter is consistent with both numerical simulations \citep{Bode2012} and dynamical mass measurements of clusters \citep{Sifon2013}. The distribution of SZ derived cluster masses ($M_{\rm SZ}$) for our sample is shown in the lower panel of Fig. \ref{fig:redshift}. We note that the values of $M_{\rm SZ}$ used in this analysis include no previous $(1-b_{\rm SZ})$ calibration.

\subsection{Weak Gravitational Lensing from the Kilo Degree Survey}
KiDS is an optical survey utilising the OmegaCAM CCD mosaic camera on the VLT Survey Telescope in Chile \citep{dejong2015}. For this analysis we are using the most recent KiDS data set, KiDS-1000, first presented in \citet{Kuijken2019}\footnote{https://kids.strw.leidenuniv.nl/DR4/index.php}. This catalogue includes 30 million galaxies across two areas of sky based around the GAMA spectroscopic survey. KiDS-1000 was observed in 4 bands, $\{u,g,r,i\}$ with an additional five bands $\{Z, Y, J, H, K_s\}$ observed by the VISTA Kilo-degree Infrared Galaxy survey \citep[VIKING,][]{Edge2013}, making KiDS a deep and wide nine-band imaging data set \citep{Wright2019}. KiDS-1000 includes 1006 deg$^2$ of imaging with galaxy lensing measurements and accurately calibrated photometric redshifts to $z_{\rm B}<1.2$ \citep{Wright2020}. The redshift distributions we use in this analysis are calibrated following the methods adopted in the KiDS-1000 cosmology analyses \citep{Asgari2020,Heymans2021} which use a self-organising map (SOM). This approach removes source galaxies from our sample because their redshift could not be accurately calibrated with spectroscopic data sets, resulting from incompleteness in the spectroscopic sample \citep{Wright2020, Hildebrandt2020}. We select source galaxies by their photometric redshift estimate, $z_{\rm B}$, which is derived from the nine-band imaging using the Bayesian Photometric Redshifts code \citep[BPZ,][]{Benitez2000}. For each cluster lensing measurement, source galaxies are selected to have $z_{\rm B}>z_{\rm l}+0.2$, where $z_{\rm l}$ is the redshift of the lens. This selection aims to eliminate unlensed galaxies from our source galaxy sample, although correction factors are still required (discussed in Sect. \ref{sec:DS}). A redshift distribution calibrated by the SOM, based on this $z_{\rm B}$ redshift selection, is produced for each cluster. The more selective nature of the SOM method has reduced the galaxy number density and therefore increased the statistical error over the same area, however, the systematic error associated with the photometric redshift distribution has decreased. Source galaxy shape measurements are computed with the {\it lens}fit pipeline \citep{Miller2013}, which is calibrated with simulations described in \citet{Kannawadi2019} and presented in \citet{Giblin2020}. KiDS-1000 has an effective number density of 6.17 arcminutes$^{-2}$, in the photometric redshift range $(0.1 < z_{\rm B} < 1.2)$, and the variance of the ellipticities is on average 0.265 \citep[see table 1,][]{Giblin2020}. For more technical details on the survey see \citet{Kuijken2019}. The footprint of the observed KiDS region is shown in Fig. \ref{fig:cluster_plot}.

\subsection{The Excess Surface Density}
\label{sec:DS}
We summarise the formalism for estimating the weak gravitational lensing signal around a stack of clusters, following \cite{BartelmannSchneider2001}. Considering an isolated cluster, weak lensing introduces a systematic tangential alignment of the images of the source galaxies relative to the lens.  We use the average tangential distortion, \(\gamma_{\rm t}\), to quantify the lensing signal. This is calculated from the two components of shear, $[\gamma_1,\gamma_2]$,
\begin{equation}
\begin{pmatrix}
\gamma_{\rm t} \\
\gamma_{\times} \\
\end{pmatrix}
=
\begin{pmatrix}
-\cos2\phi & -\sin2\phi \\
\sin2\phi & -\cos2\phi \\
\end{pmatrix}
\begin{pmatrix}
\gamma_1 \\
\gamma_2 \\
\end{pmatrix}
\, ,
\end{equation}
where, in a flat sky approximation, $\phi$ is the angle measured from a line of constant declination to a given source galaxy relative to the centre of the lens and $\gamma_{\times}$ is the shear component along the \(45^\circ\) rotated direction. In practice, \(\gamma_{\rm t}\) and \(\gamma_{\times}\) are both averaged over all angles so that a gravitational lensing signal can be detected from the average \(\gamma_{\rm t}\), and the average $\gamma_{\times}$ is a useful control statistic that should be consistent with zero for a signal solely caused by gravitational lensing. 

Since the tangential distortion we observe is due to intervening matter along the line of sight, the average $\gamma_{\rm t}$ of source galaxies, at a projected comoving separation $R$ from the cluster lens can be expressed in terms of the excess surface density, $\Delta \Sigma$,
\begin{equation}
\hat{\gamma}_{\rm t} (R) = \frac{\Delta \Sigma (R)}{\Sigma_{\rm c}(\chi_{\rm l},\chi_{\rm s})} \, ,
\label{avg_gt}
\end{equation}
where $\Sigma_{\rm c}$ is the comoving critical surface density, written in terms of comoving distances, given by
\begin{equation}
\Sigma_{\rm c} = \frac{c^2}{4\pi G}\frac{\chi_{\rm s}}{(\chi_{\rm s}-\chi_{\rm l}) \, \chi_{\rm l}} \frac{1}{(1+z_{\rm l})} \, ,
\label{sigma_crit}
\end{equation}
for spatially flat models only and $\chi$ is the comoving distance to the lens $\chi_{\rm l}$ and the source $\chi_{\rm s}$ and $z_{\rm l}$ is the lens redshift.  The excess surface density (ESD) is defined as
\begin{equation}
\Delta \Sigma (R) \equiv \overline{\Sigma} (<R) - \Sigma(R) \, ,
\label{eq:deltasigma_def}
\end{equation}
with the average mass density within a radius $R$ given by
\begin{equation}
\overline{\Sigma} (<R) \equiv \frac{2}{R^2} \int_0^R R' \, \Sigma(R') \, \mathrm{d}R' \, .
\label{eq:avg_sigmadef}
\end{equation}
Equation (\ref{avg_gt}) can be generalised for broad lens and source redshift distributions, $n_{\rm l}$ and $n_{\rm s}$ to give
\begin{equation}
\langle \gamma_{\rm t} \rangle (R) = \int_0^{\infty} \mathrm{d}z \, n_{\rm l}(z) \,  \Delta \Sigma(R,z) \int_{z}^{\infty} \mathrm{d}z' \, n_{\rm s}(z') \, \frac{1}{\Sigma_{\rm c} (z,z')}  \, .
\label{eqn:gamma}
\end{equation}
For a narrow lens and source redshift distribution, $n_{\rm l}(z) = \delta_{\rm D} (z-z_{\rm l})$ and $n_{\rm s}(z')= \delta_{\rm D}(z'-z_{\rm s})$, so that the above equation reduces to Eq. (\ref{avg_gt}). 
Eq. (\ref{eqn:gamma}) cannot be solved for in closed form, but for a narrow lens redshift distribution, averaging over a background source redshift distribution, $n(z_{\rm s})$, we find
\begin{equation}
\Delta \Sigma(R,z_{\rm l}) \approx \langle \gamma_{\rm t}\rangle(R) \left[ \int_{z_{\rm l}}^{\infty} \mathrm{d}z_{\rm s} \,n(z_{\rm s}) \, \frac{1}{\Sigma_{\rm c}(z_{\rm l},z_{\rm s})} \right]^{-1} \, .
\label{eq:deltaSigma}
\end{equation}
This equation can be re-written as
\begin{equation}
\Delta \Sigma (R,z_{\rm l}) = \frac{ \langle \gamma_{\rm t}\rangle (R) }{\overline{\Sigma^{-1}_{\rm c}}} \, ,
\end{equation}
where, for a given data set, the average inverse critical surface density, $\overline{\Sigma^{-1}_c}$, is defined as
\begin{equation}
\begin{aligned}
\overline{\Sigma^{-1}_{\rm c,l}} {} & \equiv \int_{z_{\rm l}}^{\infty} \mathrm{d}z_{\rm s} \, n(z_{\rm s}) \, \Sigma_{\rm c}^{-1} (z_{\rm l},z_{\rm s}) \\
 & = \frac{4 \pi G}{c^2} \, (1+z_{\rm l}) \, \chi_{\rm l} \int_{z_{\rm l}}^{\infty} \mathrm{d}z_{\rm s} \, n(z_{\rm s}) \, \left(1- \frac{\chi_{\rm l}}{\chi_{\rm s}} \right).
\end{aligned}
\end{equation}
From the equations derived above we can write an estimator for the ESD as
\begin{equation}
    \Delta \Sigma_i = \frac{\sum_{\rm l} \sum_{\rm s} w_{\rm ls} \epsilon_{\rm t,s} \left(\overline{\Sigma^{-1}_{\rm c,l}}\right)^{-1}\Delta_{{\rm ls,}i}}{\sum_{\rm l} \sum_{\rm s} w_{\rm ls} \Delta_{{\rm ls,}i}} \, ,
    \label{eqn:ESDest}
\end{equation}
where $\Delta_{{\rm ls,}_i}$ is a bin selector function which is defined to be unity if the radial separation between a foreground lens ${\rm l}$ and background source ${\rm s}$, lies within a bin centred on $R_i$. The tangential ellipticity of source galaxy ${\rm s}$ projected on to an axis perpendicular to the line between the lens and source is given by $\epsilon_{\rm t,s}$ and the weight, $w_{\rm ls}$ assigned to a given source galaxy, ${\rm s}$, behind a lens, ${\rm l}$,  is 
\begin{equation}
    w_{\rm ls} = w_{\rm s}\left( \overline{\Sigma^{-1}_{\rm c,l}}\right)^2 \, .
    \label{ref:weights}
\end{equation}
Since our shape measurement pipeline uses \textit{lens}fit, each source galaxy has a corresponding weight which is approximately an inverse variance weighting that accounts for total shape noise, $w_{\rm s}^{-1} \sim \sigma_e^2 + \sigma_{\rm rms}^2$, where $\sigma_{\rm rms}$ is the intrinsic shape dispersion, and $\sigma_e$ is the shape measurement error.

Due to the uncertainty associated with the photometric redshifts of the source galaxies, there will be some source galaxies that are a cluster member or a foreground galaxy. This leads to a dilution of the lensing signal and therefore a mass estimate that is biased low. This effect can be corrected for by making the same ESD measurements around 100 random positions, denoted with the subscript ${\rm r}$, that are unclustered but still have the same selection function as the lensing clusters. This approach was first shown for the tangential shear estimator \citep{Mandelbaum2006} but can analogously be applied to the ESD \citep{Singh2017} as
\begin{equation}
\Delta \Sigma_i = \frac{\sum_{\rm l} \sum_{\rm s} w_{\rm ls} \epsilon_{\rm t,ls} \left(\overline{\Sigma^{-1}_{\rm c,l}}\right)^{-1} \Delta_{{\rm ls,}i}}{\sum_{\rm r} \sum_{\rm s} w_{\rm rs}  \Delta_{{\rm rs,}i}} \left( \frac{N_{\rm r}}{N_{\rm l}} \right) - \frac{\sum_{\rm r} \sum_{\rm s} w_{\rm rs} \epsilon_{\rm t,rs} \left(\overline{\Sigma^{-1}_{\rm c,r}}\right) \Delta_{{\rm rs,}i}}{\sum_{\rm r} \sum_{\rm s} w_{\rm rs} \Delta_{{\rm rs,}i}} \, .
\label{eq:ggl_estimator}
\end{equation}
Here $N_{\rm l}$ and $N_{\rm r}$ are the number of lenses and random positions respectively within radial bin $ \Delta_{{\rm ls,}_i}$. 
Normalising with random lenses produces an unbiased estimate of the ESD unaffected by any source-lens clustering, compared to Eq. (\ref{eqn:ESDest}). This is often referred to as the `boost' factor \citep{Sheldon2004} which is given by 
\begin{equation}
    B_i = \frac{N_{\rm r}}{N_{\rm l}}\frac{\sum_{\rm l} \sum_{\rm s} w_{\rm ls}  \Delta_{{\rm ls,}i}}{\sum_{\rm r} \sum_{\rm s} w_{\rm rs} \Delta_{{\rm rs,}_i}} \, ,
\end{equation}
for the case of lenses and random positions having no additional weighting applied. 
In our analysis we ensure our random sample has the same area and redshift properties as our cluster sample.

We apply an additional correction factor to account for multiplicative bias shear calibration which accounts for uncertainty in the shape measurement calibration \citep{Kannawadi2019} and multiplies the ESD estimator defined in Eq. \ref{eq:ggl_estimator} as $\frac{1}{(1+\overline{m}_i)}\Delta \Sigma_{i} \,$
for
\begin{equation}
    \overline{m}_i = \frac{\sum_{\rm l}\sum_{\rm s}w_{\rm ls}m_{\rm s}\Delta_{ls, i}}{\sum_{\rm l}\sum_{\rm s}w_{\rm ls}\Delta_{ls, i}} \, .
\end{equation}
Here $m_{\rm s}$ is given by the multiplicative bias value of the tomographic bin which the source $\rm s$ falls into based on its $z_{\rm B}$ value. The value of the multiplicative bias per tomographic bin are described in \cite{Asgari2020} and were determined in \cite{Kannawadi2019}. 

For the stacked ESD measurement we estimate the covariance matrix following the same procedure as \citet{Miyatake2019} as
\begin{equation}
    \tens{C} = \tens{C^{\rm stat} + C^{\rm lss} + C^{\rm int}}  \, ,
\end{equation}
where $\tens{C}^{\rm stat}$ accounts for the statistical uncertainty due to galaxy shapes, $\tens{C}^{\rm lss}$ is the covariance due to projection effects from uncorrelated large-scale structure \citep{Hoekstra2001} and $\tens{C}^{\rm int}$ corresponds to the intrinsic variations due to halo triaxiality and correlated halos. The contribution to the covariance from shape noise is diagonal by definition and is given by
\begin{equation}
    \sigma^2_{{\rm stat},i} =  \sigma_e^2  \left[ \frac{\sum_{\rm ls} \left(w_{\rm ls}\overline{\Sigma^{-1}_{\rm c,l}}\right)^2 \Delta_{{\rm ls,}i}}{\left(\sum_{\rm rs} w_{\rm rs} \overline{\Sigma^{-1}_{\rm c,r}}^2  \Delta_{{\rm rs,}i} \right)^2}  \left( \frac{N_{\rm r}}{N_{\rm l}} \right)^2 +  \frac{\sum_{\rm rs} \left(w_{\rm ls}\overline{\Sigma^{-1}_{\rm c,l}}\right)^2  \Delta_{{\rm rs,}i}}{\left(\sum_{\rm rs} w_{\rm rs} \overline{\Sigma^{-1}_{\rm c,r}}^2  \Delta_{{\rm rs,}i}\right)^2} \right]  \, ,
\end{equation}
for the estimator given in Eq. (\ref{eq:ggl_estimator}). The large-scale structure and intrinsic parts of the covariance are defined in Equations A1 and A7 in \citet{Miyatake2019}. 

\section{Results}
\label{sec:result_stack}
\begin{figure}
	\includegraphics[width=\columnwidth]{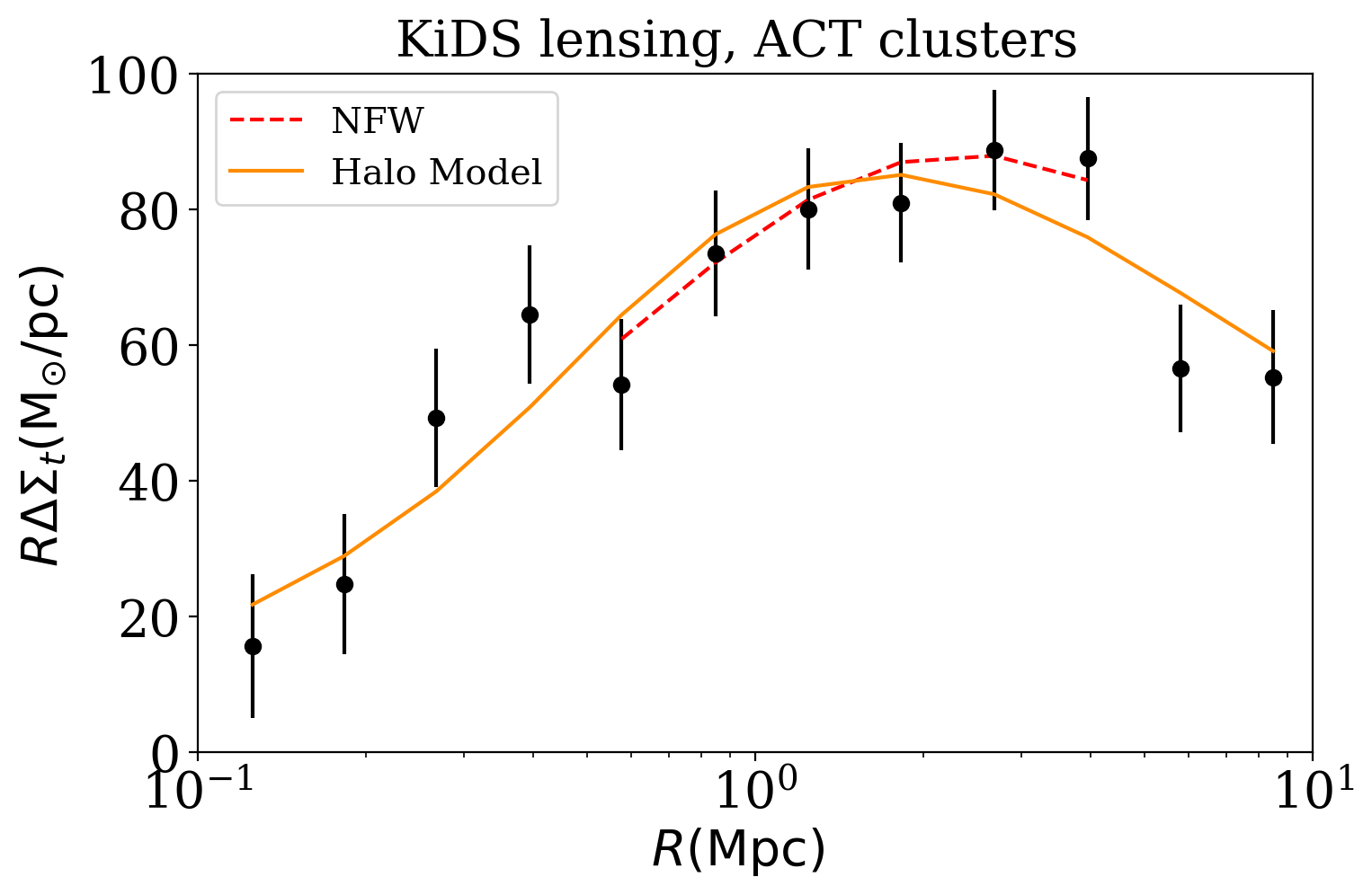}
    \caption{The ESD profile ($\Delta\Sigma$) multiplied by the radius $R$ as a function of radius $R$ for the complete stack of 157 SZ-detected clusters. The signal is detected at 25$\sigma$ significance. The best-fitting model is shown using a single NFW, shown in red, and the halo model, shown in orange. A limited range of scales is used for the NFW model to avoid the impact of miscentering at small scales and the contribution from large-scale structure at scales greater than 5~Mpc. Both models give good fits to the data.
    }
    \label{fig:ESD_total}
\end{figure}

In our fiducial analysis we measure the ESD, $\Delta \Sigma$, `stacking' all $157$ ACT-detected clusters, as given by Eq. \ref{eqn:ESDest}. The lensing signal is detected with a signal-to-noise of 25 in the range 0.1~Mpc to 10~Mpc from the cluster centres, and the ESD profile is shown in Fig. \ref{fig:ESD_total}. The detection significance is computed as $\sqrt{\sum_i \Delta\Sigma_i^2 / \sigma_i^2}$.
To test the robustness of this result we investigate the impact of a set of systematic effects, detailed in appendix \ref{app}.

\subsection{Modelling the Stacked ESD}

We estimate the weak lensing mass, $M_{\mathrm{WL}}$, of the stacked clusters using two models: a single NFW halo \citep{Navarro1997} and a full halo model \citep{Peacock2000,Seljak2000}. 
The single NFW fit assumes that the measured lensing signal is only due to the cluster. The halo model also assumes that the central cluster halo is described by an NFW but additionally accounts for line-of-sight lensing.
Furthermore, the halo model method is superior for modelling the ESD signal, as the ACT selection function can be included and we also use a modified NFW profile to account for mis-centering at small scales. 
We include the NFW fit to compare our results with previous analyses which have utilised a similar modelling approach. 

The signal we observe is actually the reduced shear $g = \gamma/(1-\kappa)$, where $\kappa$ is the convergence which accounts for the change in size of an object due to gravitational lensing. At large scales $\kappa$ is very small and so the true shear, $\gamma \approx g$. At small scales, however, $\kappa$ is larger and the equivalence of $g$ and $\gamma$ no longer holds.
Given that we use data at small scales, we apply a correction to our model for the ESD signal to account for reduced shear following \citet{schrabback2018}. 

This factor simply multiplies the model for the ESD as $f_{\rm reduce} \, \Delta \Sigma^{\rm model}$ and is given for a stack of clusters as
\begin{equation}
    f_{\rm reduce} = \left[ 1 + \left( \frac{\langle \beta_s^2 \rangle}{\langle \beta_s \rangle ^2} - 1\right) \langle \beta_s \rangle \kappa_{\infty}^{\rm model}\right] \frac{1}{(1-\langle \beta_s \rangle \kappa_{\infty}^{\rm model})} \, ,
\end{equation}
where $\kappa_{\infty}^{\rm model}$ is the model convergence computed at infinite redshift and the shape weights are accounted for as
\begin{equation}
    \langle \beta_s \rangle = \frac{\sum \beta_{\rm l} w_{\rm l}}{\sum w_{\rm l}} \, , \, \langle \beta_s^2 \rangle = \frac{\sum \beta^2_{\rm l} w_{\rm l}}{\sum w_{\rm l}} \, ,
\end{equation}
where $\beta_{\rm l}$ is computed per lens for the estimated source redshift distribution as
\begin{equation}
    \beta_{\rm l} = \frac{\chi_{\infty}}{(\chi_{\infty} - \chi_{\rm l})} \int_{z_{\rm l}}^{\infty} \mathrm{d}z_{\rm s} \, n(z_{\rm s}) \frac{(\chi_{\rm s} - \chi_{\rm l})}{\chi_{\rm s}} \, .
\end{equation}
The $\frac{\langle \beta_s^2 \rangle }{ \langle \beta_s \rangle ^2}$ in the correction term accounts for the width in the redshift distribution \citep{Hoekstra2000}. For our sample this correction factor is $\sim $1.3 at the smallest scales, which correspond to $1 \sigma$ shift in the ESD. For $R > 0.5 {\rm Mpc}$ the correction factor decreases to less than 1.05.

\subsubsection{Single NFW}
We estimate the average cluster mass, $M_{\mathrm{WL}}$, by assuming an NFW density profile \citep{Navarro1997}
and calculating the expected ESD using \cite{Wright2000}, summarised here. The NFW profile is given by
\begin{equation}
	\rho(r) = \frac{\delta_c \rho_{\rm c}}{(r/r_{\rm s})(1+r/r_{\rm s})^2} \, ,
\label{eq:NFW}
\end{equation}
where the characteristic overdensity, $\delta_c$, is related to the NFW concentration parameter, $c_{500}$ via
\begin{equation}
    \delta_c = \frac{500}{c_{500}} \frac{c_{500}^3}{[\ln(1+c_{500})-c_{500}/(1+c_{500})]} \, .
\label{eq:deltac}
\end{equation}
In Eq. ~\ref{eq:NFW}, the scale radius, $r_{\rm s} = r_{\rm 500c}/c_{500}$, is the cluster's characteristic radius. We use the $r_{500c}$ radius to be consistent with the SZ measurements which also use this definition. The mass within a radius of $r_{\rm 500c}$, is given by
\begin{equation}
    M_{\rm 500c} \equiv M(r_{\rm 500c}) = 500 \rho_{\rm c} \, \frac{4\pi}{3} \, r_{\rm 500c}^3 \, .
\end{equation}
Within this model we fit the mass, $M_{\rm 500c}^{\mathrm{WL}}$, and the concentration, $c_{500}$, using wide flat priors and take the average cluster redshift of $z=0.36$, which is calculated by weighting the clusters using the weights defined in Eq. (\ref{ref:weights}). Fitting for the concentration, rather than assuming a mass-concentration relation which has been calibrated on dark matter-only simulations, has been shown to capture the change in the ESD profile due to baryon feedback which effectively redistributes material from the centre to the outer regions of the cluster. \citet{Debackere2021} showed that when assuming a mass-concentration relation, the mass inferred from weak lensing can be overestimated by 10\% at the smallest scales and underestimated by around 5\% at the largest scales considered in our analysis. Allowing the concentration and mass to vary removes the bias at large scales and reduces the bias at small scales to less than 5\% for the largest halos, which are most affected, and to $\sim$1\% for the halos with a mass comparable to those considered in this work. Thus this approach for mitigating the impact of baryons is sufficient in decreasing the bias in the mass estimate. \citet{Debackere2021} demonstrated however that when performing a cosmological analysis the bias in cosmological parameters increases due to a mis-match in the mass definition when the mass function is obtained from dark matter-only simulations. As a result the inferred bias cannot be directly applied as a calibration in a cosmological analysis.

The best-fitting NFW model is shown in Fig. \ref{fig:ESD_total}, which has a p-value of 0.81. We only use the range $(0.5<R<5)~\rm{Mpc}$ to avoid any miscentering biases at small scales (see appendix \ref{app}), and to avoid including the potentially significant signal from the large-scale structure at scales greater than $5 \, {\rm Mpc} \,$ \citep{Applegate2014}. This upper limit was chosen apriori and the lower limit was informed by the analysis presented in appendix \ref{app} although this is consistent with the ranges used in previous cluster weak lensing analyses. We compute the average weak lensing mass to be
$M_{\rm 500c}^{\mathrm{NFW}}= (4.9\pm 0.4) \times 10^{14}\mathrm{M}_{\odot}$.

\subsubsection{Halo Model Method}
Cluster lensing can be understood within the framework of the halo model following \citet{Seljak2000,Peacock2000,Cooray2002}. The projected surface density is completely specified by the cluster--dark matter cross correlation, $\xi_{\rm c,dm}$, which can be computed from a given halo occupation model,
\begin{equation}
	\Sigma(R) = 2 \bar{\rho} \int_R^{\infty}  \xi_{\rm c,dm}(r) \frac{r \, \mathrm{d}r}{\sqrt{r^2-R^2}} \, ,
    \label{eq:x-cor}
\end{equation}
where $r$ is the 3D comoving distance $r^2 = R^2 + (\chi - \chi_{\rm l})^2$, for a flat Universe. The cluster--dark matter cross correlation power spectrum, at a wavenumber $k$, is given in terms of $\xi_{\rm c,dm}$ as
\begin{equation}
	P_{\rm c,dm}(k) = 4\pi \int_0^\infty \xi_{\rm c,dm} (r) \frac{\sin (kr)}{kr} r^2 \mathrm{d}r \, .
    \label{eq:pwr_spec}
\end{equation}
The power spectrum of the cluster--dark matter cross correlation can be split into two parts:
\begin{equation}
\begin{split}
	P_{\rm c,dm}(k) = P_{\rm c,dm}^{\rm 1h}(k)  + P_{\rm c,dm}^{\rm 2h}(k) \, ,
\end{split}
    \label{eq:halo_model}
\end{equation}
with the `1--halo term', which includes the non--linear regime at small scales, and the `2--halo term' which describes the clustering between halos, which is dominant on large scales.

The `1--halo' term, $P_{\rm c,dm}^{\rm 1h}(k)$, describes the dark matter distribution inside halos hosting clusters. For a single cluster, this is 
\begin{equation}
	P_{\rm c,dm}^{\rm 1h} (k) = \frac{1}{\bar{\rho}} u_{\rm c} (k|M) \, ,
\end{equation}
where $u_{\rm c}(k|M)$ is the Fourier transform of the cluster density profile $\rho(r|M)$, which we assume to be described by an NFW profile. 

Since we are measuring a stack of clusters, the `1--halo' term becomes 
\begin{equation}
	P_{\rm c,dm}^{\rm 1h}(k) = \frac{1}{\bar{\rho}} \int_0^\infty \mathcal{P}(M) u_{\rm c}(k|M) \mathrm{d}M \, ,
    \label{eq:1halo_def}
\end{equation}
where $\mathcal{P}(M)$ is the probability that a cluster within an SZ sample resides in a halo of mass $M$, which is given
\begin{equation}
	\mathcal{P}(M) \, \mathrm{d}M = \frac{\left\langle N_{\rm c}|M\right\rangle \frac{\mathrm{d}n(M)}{\mathrm{d}M}}{\bar{n}_{\rm c}} \mathrm{d}M\, ,
    \label{eq:bayes}
\end{equation}
where $\left\langle N_c|M\right\rangle$ is the average number of clusters in the SZ selection with a halo of mass $M$, $\frac{\mathrm{d}n(M)}{\mathrm{d}M}$ is the halo mass function and $\bar{n}_{\rm c} = \int \langle N_{\rm c}|M \rangle \frac{\mathrm{d}n(M)}{\mathrm{d}{M}} \mathrm{d}M$ is the comoving number density of clusters in the SZ selection. Equation (\ref{eq:1halo_def}) can therefore be re-written as:
\begin{equation}
	P_{\rm c,dm}^{\rm 1h} (k) = \frac{1}{\bar{\rho}\bar{n}_{\rm c}} \int_0^\infty \mathrm{d}M \, \frac{\mathrm{d}n(M)}{\mathrm{d}M}u_{\rm c}(k|M) \, \langle N_{\rm c}|M \rangle \, .
    \label{eq:1halo}
\end{equation}
This `1--halo' term is constructed so that mis-centering of clusters can be accounted for in the cluster density profile, via the $u_c(k|M)$ term defined as
\begin{equation}
	u_{\rm c}(k|M) = u_{\rm dm}(k|M) (1-p_{\rm off} + p_{\rm off}e^{-0.5k^2(r_{\rm s} f_{\mathrm{off}})^2}) \, ,
    \label{eq:mis-center}
\end{equation}
where, $p_{\rm off}$ and $f_{\rm off}$ is the probability of mis-centering and the size of the offset, as fraction of $r_{\rm s}$,  respectively. 

The `2--halo' term, denoted by $P_{\rm c,dm}^{\rm 2h}(k)$, describes the correlation between clusters and dark matter particles belonging to separate haloes, and is given by 
\begin{equation}
\begin{split}
	P_{\rm c,dm}^{\rm 2h}(k) = \frac{P_{\rm dm}(k)}{\bar{\rho}}\int_0^\infty & \mathcal{P} (M) b_{\rm h}(M) \,\mathrm{d}M \\
    & \int_0^\infty u_{\rm dm}(k|M')b_{\rm h}(M')\frac{\mathrm{d}n(M')}{\mathrm{d}M'} \, \mathrm{d}M' \, .
\end{split}
\label{eq:2term}
\end{equation}
Here $P_{\rm dm}$ is the matter power spectrum, which we assume to be linear, and $b_{\rm h}(M)$ is the halo bias function. The halo bias function describes how halos of mass $M$ are biased with respect to the overall dark matter distribution. Using Eq. (\ref{eq:bayes}), we can re--write Eq. (\ref{eq:2term}) as 
\begin{equation}
\begin{split}
	P_{\rm c,dm}^{\rm 2h}(k) = \frac{P_{\rm dm}(k)}{\bar{\rho}\bar{n}_{\rm c}}\int_0^\infty & \langle N_{\rm c}|M \rangle b(M) \frac{\mathrm{d}n(M)}{\mathrm{d}{M}} \, \mathrm{d}{M} \\
    & \int_0^\infty u_{\rm dm}(k|M')b(M')\frac{\mathrm{d}n(M')}{\mathrm{d}{M'}} \, \mathrm{d}{M'} \, .
\end{split}
\end{equation}
We adopt the halo bias model from \citet{Tinker2010} and include an amplitude $A_{\rm b}$ that multiplies the halo bias as $b(M) = A_{\rm b}b_{\rm h}(M)$, which we marginalise over in our halo model analysis.
For the halo occupation statistics that are required to calculate $\langle N_{\rm c}|M \rangle$, we assume a simple power-law dependence between halo mass $M$ and $M_{\mathrm{SZ}}$
\begin{equation}
	M_{\mathrm{SZ}} = A M^{\beta}  \, ,
\end{equation}
to calculate the average number of clusters. Given the limited signal-to-noise ratio of the data we fix the mass-dependent exponent, $\beta$, equal to 1, so that we can focus on $A$, which quantifies the constant mass bias between the SZ mass and the true mass. For our sample of clusters, the average number of clusters is then given by,
\begin{equation}
	\langle N_c|M \rangle = \frac{1}{\sqrt{2 \pi} \sigma_c}\int_0^{\infty}  \frac{S(M_{\rm SZ})}{M_{\rm SZ}} {\rm e}^{ \left[ \frac{-(\ln({M_{\mathrm{SZ}}/M}) - \ln{A})^2}{2\sigma_c^2}\right]} \, \mathrm{d}{M_{\rm SZ}} \, ,
\label{eq:occupation}
\end{equation}
where we have assumed a log-normal scatter, $\sigma_c$, and $S(M_{\rm SZ})$ is the SZ-experiment selection function, which can be computed for a given range in SZ signal-to-noise. We fix the SZ mass scatter to $\sigma_c = 0.2$, firstly because lensing does not constrain this well and secondly because this is the same scatter taken to correct the Eddington bias selection effect for the SZ masses derived for the ACT clusters \citep{Hilton2020}, based on the level of scatter seen in numerical simulations and dynamical mass measurements. We check that the chosen value for the scatter does not significantly impact our estimate of $(1-b_{\rm SZ})$ by repeating our analysis with $\sigma_c=0.001$ and $\sigma_c=0.5$. Our inferred value for the mass changes by less than $0.2 \sigma$ which has a 2\% effect on $(1-b_{\rm SZ})$.

In this model we fit for the following set of parameters: $\lbrace c, \ln{A}, A_b, p_{\mathrm{off}}, f_{\mathrm{off}} \rbrace$.
The average halo mass, which is the main parameter we are interested in, is a derived parameter, given by
\begin{equation}
	M_{\rm 500c}^{\mathrm{HM}} \equiv M_{\rm h} = \frac{1}{\bar{n}_{\rm c}} \int \langle N_{\rm c}|M \rangle M  \frac{\mathrm{d}n(M)}{\mathrm{d}{M}} \, \mathrm{d}M \, .
    \label{eq:halo_mass}
\end{equation}
Our halo model framework described above is built from the halo model modules in the public Core Cosmology Library \citep{CCL}.

We use Bayesian inference to recover the posterior probability distribution of these five parameters using {\sc emcee}, the MCMC Python package \citep{emcee}. We use wide flat priors for all parameters $(c\in[0,10], \, \ln{A}\in[-5,5], \, A_b\in[0,10], \, p_{\rm off}\in[0,1], \, f_{\rm off}\in[0,1])$ and assume a Gaussian likelihood. We fix the cosmological parameters to \citet{Planck2018}, which are in excellent agreement with the recent CMB lensing results from ACT \citep{Madhavacheril2023}, for computing $P_{\rm dm}$. Assuming a value for $\sigma_8$ which has been determined at a similar redshift to our lensing measurements is more robust as this is not dependent on assuming structure growth. We find our results are insensitive to this choice however, by repeating our analysis assuming cosmological parameters determined from KiDS cosmic shear measurements \citep{Heymans2021}, which are in $\sim 2 \sigma$ tension with \textit{Planck}, where we recover the same estimate of the cluster mass. Furthermore, if we marginalise over $\sigma_8$, $\Omega_{\rm m}$ and $h$ in our fit we recover the same value for the inferred mass but our error on the mass increases by 50\%.  As we have a mis-centering term and a 2-halo term in this model we use the full range of scales from $(0.1 - 10)\,{\rm Mpc}$. We check that our mass estimate is robust to the choice of scales included in the analysis by re-performing our fits with the limited range of scales used in the single NFW halo fit, the mass estimate remains the same but the error increases by 20\%. The halo model is evaluated at a median redshift of $z=0.36$, calculated using the same weighting defined in Eq. \ref{ref:weights}. From our MCMC sampling analysis, we infer the average weak lensing mass estimate to be $M^{\mathrm{HM}}_{\rm 500c}= (4.9 \pm 0.4) \times 10^{14}\mathrm{M}_{\odot}$. Our best-fitting halo model is shown in Fig. \ref{fig:ESD_total} and has a p-value of 0.28. 

\begin{figure}[h!]
	\includegraphics[width=\columnwidth]{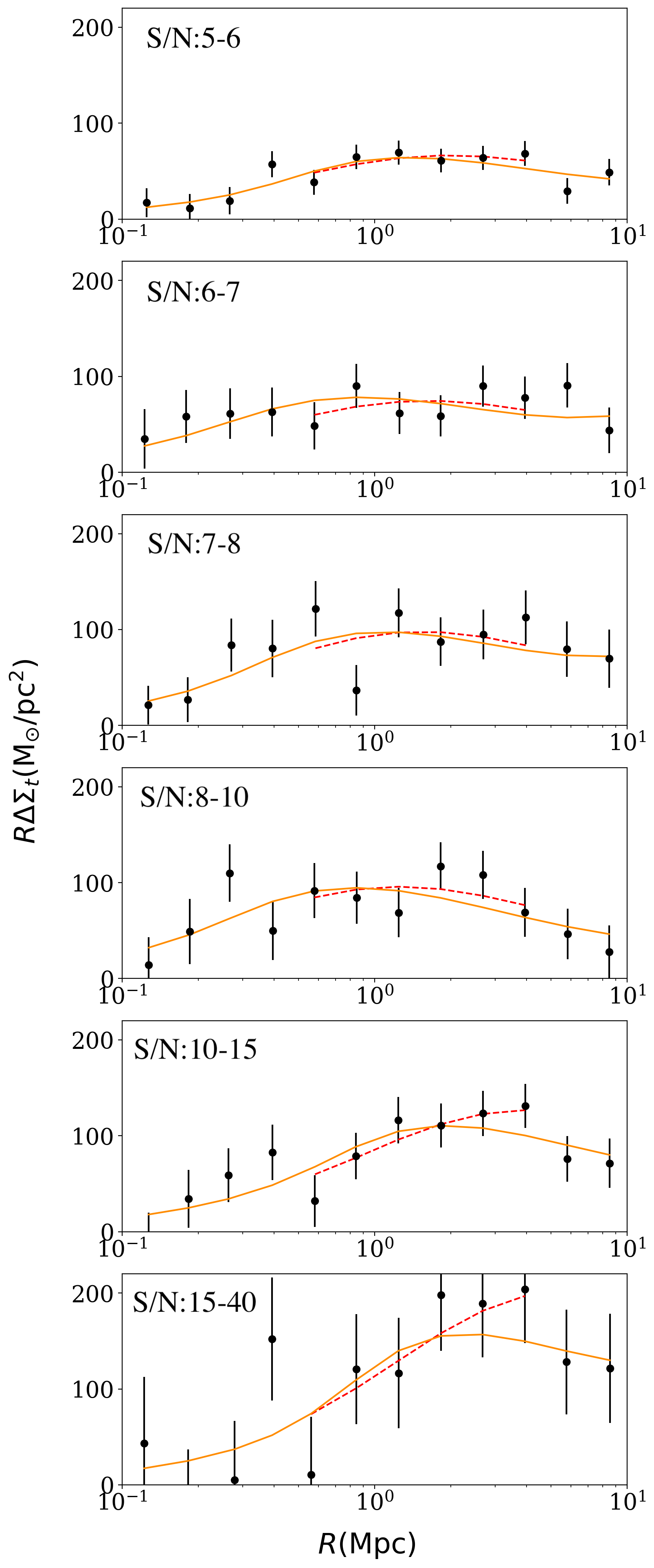}
    \caption{The ESD profile ($\Delta\Sigma$) as a function of radius $R$, for each sub-sample of clusters. The best fit lines are also plotted shown in red for the single NFW and orange for the halo model.}
    \label{fig:ESD_binned}
\end{figure}
\begin{table*}[h!]
\caption{We bin clusters based on their SZ signal-to-noise. The average SZ mass and redshift for a given bin is computed using the weights defined in Eq. (\ref{ref:weights}).}
\centering
\begin{tabular}{c c c c c c c c} 
 SZ S/N range & no. of clusters & $\overline{M}_{\mathrm{SZ}}(10^{14}\mathrm{M_{\odot}})$ & $\overline{z}$ & $M_{\mathrm{NFW}}(10^{14}\mathrm{M_{\odot}})$ & $(1-b_{\rm SZ})_{\mathrm{NFW}}$ & $M_{\mathrm{HM}}(10^{14}\mathrm{M_{\odot}})$ & $(1-b_{\rm SZ})_{\mathrm{HM}}$ \\ [0.5ex] 
 \hline
 5 -- 40 & 157 & 3.18$\pm$0.07 & 0.36 & 4.9$\pm$0.4 & 0.64$\pm$0.05 & 4.9$\pm$0.4 & 0.65$\pm$0.05 \\
 \hline
 5 -- 6 & 76 & 2.46$\pm$0.08 & 0.36 & 3.3$\pm$0.5 & 0.7$\pm$0.1 & 3.3$\pm$0.5 & 0.7$\pm$0.1 \\ 
 6 -- 7 & 26 & 2.9$\pm$0.2 & 0.38 & 4.0$\pm$0.8 & 0.7$\pm$0.2 & 3.8$\pm$0.9 & 0.7$\pm$0.2\\
 7 -- 8 & 15 & 3.3$\pm$0.2 & 0.33 & 5.7$\pm$1.1 & 0.6$\pm$0.1 & 5.3$\pm$1.2 & 0.6$\pm$0.1 \\
 8 -- 10 & 21 & 3.5$\pm$0.2 & 0.43 & 6$\pm$1 & 0.6$\pm$0.1 & 4.9$\pm$0.9 & 0.7$\pm$0.1  \\
 10 -- 15 & 16 & 4.3$\pm$0.3 & 0.31 & 8$\pm$1 & 0.6$\pm$0.1 & 7$\pm$1 & 0.6$\pm$0.1  \\ 
 15 -- 40 & 3 & 9$\pm$1 & 0.31 & 15$\pm$5 & 0.6$\pm$0.2 & 12$\pm$4& 0.7$\pm$0.2 \\ [1ex] 
 \hline
\end{tabular}
\label{tab:cluster_bins}
\end{table*}
\subsection{Splitting the cluster sample}
Given our large cluster sample and high signal-to-noise in the stacked cluster lensing measurement we split our sample to investigate the dependence of calibration bias on the cluster mass. As the inferred SZ cluster mass has a large associated uncertainty we split based on the SZ signal-to-noise, since the SZ signal is correlated with the cluster mass. The binning is defined in Table \ref{tab:cluster_bins} with the corresponding number of clusters in each bin, average SZ inferred mass and average cluster redshift.

We perform both NFW and halo model fits to each signal-to-noise bin following the same procedure as described for the total stack of clusters above. The best fit values are reported in Table \ref{tab:cluster_bins} for both models. The ESD signals for each bin with corresponding best fit models are shown in Fig. \ref{fig:ESD_binned}.

\section{Calibration bias: $(1-b_{\rm SZ})$}
\label{sec:b}
\begin{figure*}[h!]
    \includegraphics[width=\textwidth]{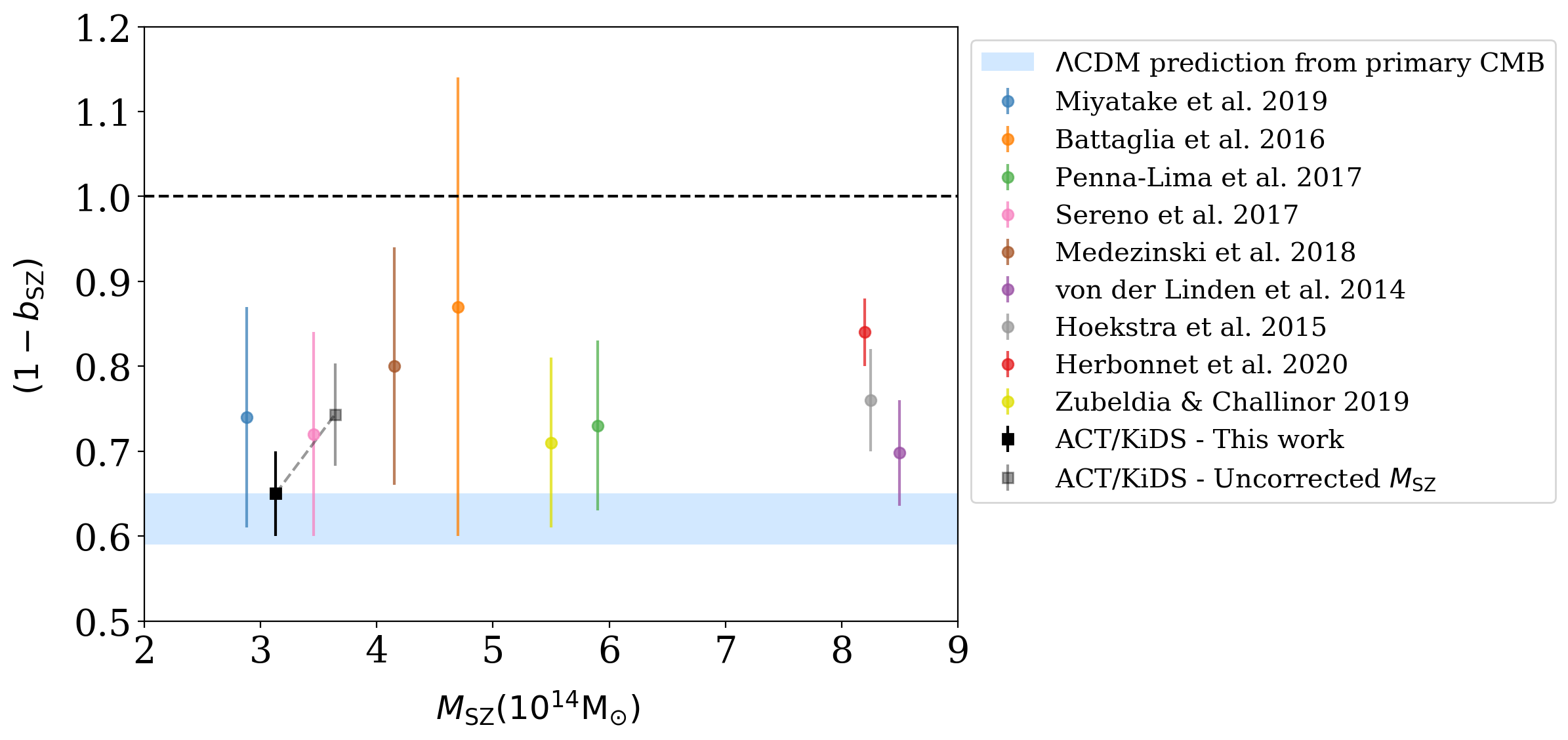}
    \caption{The estimated $(1-b_{\rm SZ})$ bias parameter from this analysis (ACT SZ-detected clusters calibrated with KiDS weak lensing), compared to other recent results. Shown on this figure at the lower mass range, the analyses of ACT and \textit{Planck} clusters using HSC \citep{Miyatake2019,medezinski2018}, ACT/CS82 \citep{battaglia2016}, CLASH \citep{penna-lima2017}, PSZ2LenS \citep{Sereno2017}, WtG \citep{vonderlinden2014b}, CCCP \citep{hoekstra2015} which has recently been updated to CCCP and MENeaCS \citep{Herbonnet2020} and \textit{Planck} clusters with \textit{Planck} CMB lensing \citep{Zubeldia2019} analyses. The blue band indicates the prediction from \textit{Planck} primary CMB measurements if the $\Lambda$CDM model is correct \citep{Planck2018}, given the SZ clusters observed by \textit{Planck}. This value is computed using a \citet{Planck2018} primary CMB best fit cosmology, however using the cosmology derived from ACT DR6 lensing, \citet{Madhavacheril2023}, would give the same result.}
    \label{fig:mass}
\end{figure*}
To compute the mass bias $(1-b_{\rm SZ})$ we combine estimates of the SZ inferred cluster masses by taking the weighted average. Here clusters are weighted in the same way as in the stacked ESD measurement described in Eq. (\ref{ref:weights}). We estimate the average SZ inferred mass to be $M^{\mathrm{SZ}}_{\rm 500c} = (3.18\pm0.07)\times10^{14}\mathrm{M}_{\odot}$.

\begin{figure}
    \includegraphics[width=\columnwidth]{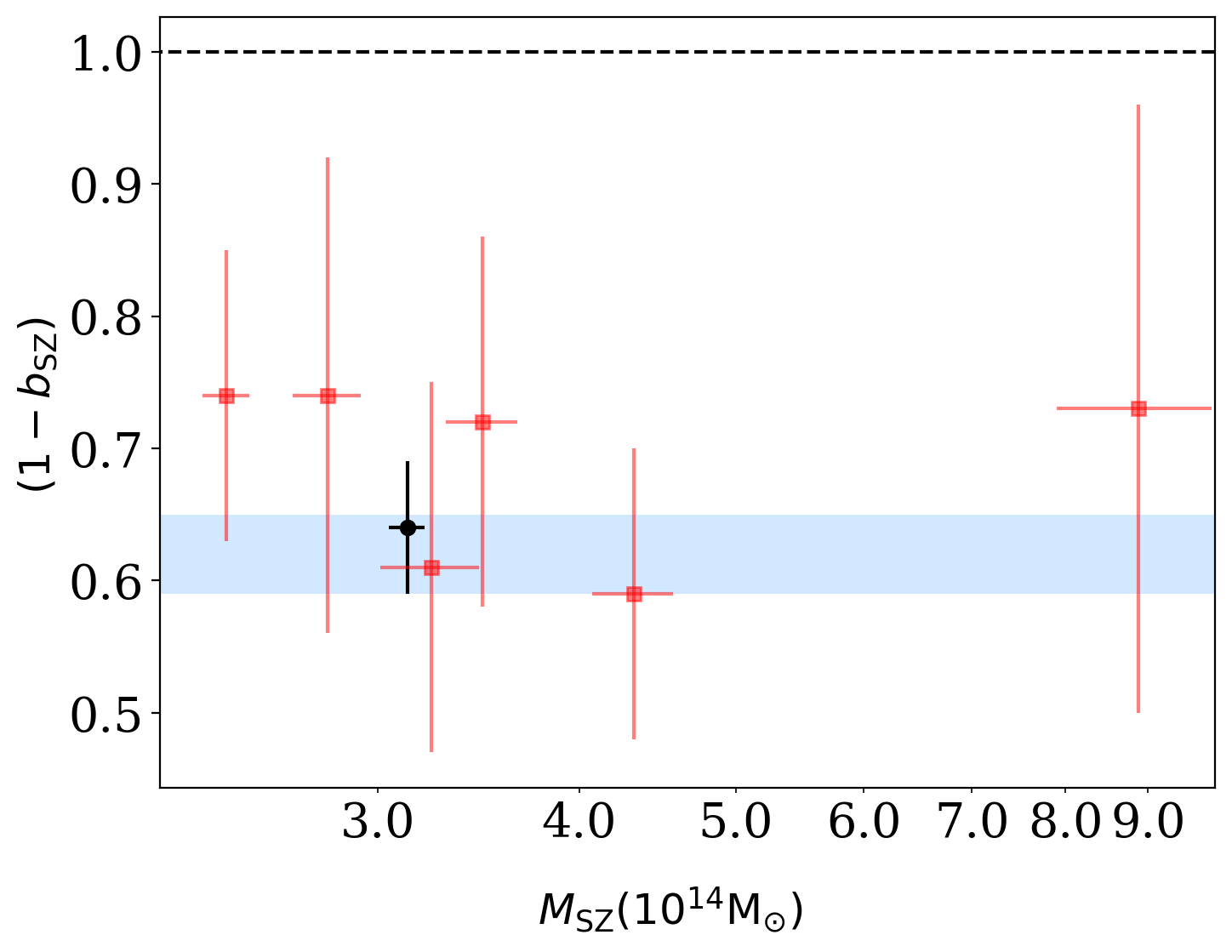}
    \caption{The estimated $(1-b_{\rm SZ})$ bias parameter from the total cluster stack shown in black, and in red for six individual SZ signal-to-noise bins. These values are reported in Table \ref{tab:cluster_bins}. The blue band indicates the prediction from \textit{Planck} primary CMB measurements if the $\Lambda$CDM model is correct \citep{Planck2018}, given the SZ clusters observed by \textit{Planck}.}
    \label{fig:mass_binned}
\end{figure}

From our halo model estimate of the weak lensing mass, we estimate $(1-b_{\rm SZ})=0.65\pm 0.05$. This is consistent with our estimate using a single NFW profile of $(1-b_{\rm SZ})=0.64\pm0.05$. The large deviation of $(1-b_{\rm SZ})$ from unity suggests that clusters are not in hydrostatic equilibrium and that baryon feedback plays an important role; this is consistent with recent results from kinetic SZ and thermal SZ analyses \citep[for example,][]{Koukoufilippas2020,Ibitoye2022}. This bias, however, cannot be solely attributed to the assumption of hydrostatic equilibrium and is dependent on other assumptions made when estimating $M_{\rm SZ}$. In Fig. \ref{fig:mass} we show our new estimate together with a compilation of measurements from other data combinations. ACT and \textit{Planck} cluster mass estimates can be directly compared since they are derived from the same SZ-mass scaling relations and pressure profiles. Our results are consistent with, though generally lower than, analyses of ACT and \textit{Planck} clusters in a similar mass range using weak galaxy lensing data from Hyper Suprime-Cam \citep[HSC; see][]{Miyatake2019,medezinski2018}, CS82 \citep{battaglia2016} and CFHTLenS/RCSLenS \citep{Sereno2017}. We also find consistency with the analysis of \textit{Planck} clusters with CLASH galaxy weak lensing \citep{penna-lima2017} and \textit{Planck} CMB lensing \citep{Zubeldia2019}. Compared to several of the earlier analyses, shown in Fig. \ref{fig:mass},  we have a slight improvement in constraining power which is due to the larger cluster sample.

Comparing to higher mass \textit{Planck} cluster analyses, WtG \citep{vonderlinden2014b}, CCCP \citep{hoekstra2015} and its more recent re-analysis CCCP/MENeaCS \citep{Herbonnet2020}, our inferred value for the mass bias is up to $3.8\sigma$ lower. \citet{battaglia2016} demonstrated, however, that adding a correction for Eddington bias (see section 3.1) applied to the SZ mass estimates, would lower the value of $(1-b_{\rm SZ})$ in both the WtG and CCCP analysis \citep[see Fig. 7 in][]{battaglia2016}. We do apply that correction in this work. In Fig. \ref{fig:mass} we show how our result would change if the correction was not applied, with the value we estimate for the mass bias increasing to $(1-b_{\rm SZ})=0.74\pm 0.06$, which is a shift of $2 \sigma$. It appears that the tension between the value for $(1-b_{\rm SZ})$ found in this work and the recent result from \citet{Herbonnet2020}, who found  $(1-b_{\rm SZ})=0.84\pm0.04$, can be explained by uncorrected Eddington bias in the latter.

We also find consistency with the mass bias required for \textit{Planck} primary CMB cosmology to be consistent with the \textit{Planck} clusters counts analysis, which is indicated by the blue band in Fig. \ref{fig:mass}. To be able to combine all the measurements shown in Fig. \ref{fig:mass}, we would need to account for covariances between different data sets since some clusters appear in more than one sample and the difference in the ACT and \textit{Planck} selection functions. 

By splitting our sample based on SZ detection signal-to-noise, a quantity correlated with the cluster mass, we investigate whether the bias is mass dependent. The results are reported in Table \ref{tab:cluster_bins} and shown in Fig. \ref{fig:mass_binned}. We fit a straight line to ${M_{\rm SZ}}$ against $(1-b_{\rm SZ})$, considering two cases: the first just fitting for the $y$-intercept (no mass dependence) and the second fitting for both the gradient and the y-intercept. We find no evidence for any dependence on mass; with a best fit gradient of $0\pm0.1$.   
We find that repeating the fit with a gradient of zero still produces an excellent fit, a p-value of 0.9, and only improves the p-value at the sub-percent level when including a mass dependence. Similar results are obtained using the best fit mass estimates from the NFW model.

\section{Conclusions}
\label{sec:conclude}
We have estimated the masses of a new sample of galaxy clusters detected using the Atacama Cosmology Telescope.  Stacking the signal from $157$ high signal-to-noise clusters, we detect the weak lensing signal at $25 \sigma$ significance. We estimate the stacked mass to be $M^{\mathrm{WL}}_{\rm 500c}= (4.9 \pm 0.4) \times 10^{14}\mathrm{M}_{\odot}$, using an analytical halo model formalism which accounts for the ACT SZ cluster selection. By comparing this gravitational lensing estimate of mass to the mass inferred from the SZ signal itself, we provide a new estimate of the calibration of the SZ Compton--$y$ parameter mass scaling relation.

When modelling the cluster lensing signal we adopt two methods, a simpler NFW and a full halo model. The NFW method only models the central halo term and we therefore limit the scales used for this fit to avoid mis-centering at small scales and large-scale structure at scales greater than $5 \, {\rm Mpc}$. The halo model method assumes an NFW profile for the central halo, however it also accounts for line-of-sight lensing, the ACT selection function and mis-centering. Both mass estimates are consistent within $1 \sigma$ of each other and have comparable constraining power, which may be initially surprising. This shows that even though we are able to include more data at large and small scales in the halo model fit, the uncertainty on centroiding and the large-scale structure means these extra scales do not help to constrain the mass. The comparable constraining power of these two methods reinforces that our scale cuts for the NFW model fit are correctly defined. 

We find our estimate of the mass calibration bias to be consistent, given the statistical uncertainty, with estimates made using other contemporary weak lensing surveys including HSC. This measurement does not indicate a departure from the predicted value of the mass bias from \textit{Planck} primary CMB measurements if the $\Lambda$CDM model is correct \citep{Planck2018}, given the SZ clusters observed by \textit{Planck}. Additionally, we find no tension with estimates of the mass bias from a higher mass sample if we do not account for Eddington bias, as first demonstrated in \citet{battaglia2016}.

From splitting our sample based on SZ detection signal-to-noise, which is correlated with the cluster mass, we investigated if the bias has a dependence on mass. Within the statistical uncertainty of our measurements we found no evidence of any dependence on mass which is consistent with the results presented in \citet{Herbonnet2020}.

\begin{acknowledgements}
The figures in this work were created with \texttt{matplotlib} \citep{matplotlib} making use of the \texttt{numpy} \citep{numpy}, \texttt{scipy} \citep{2020SciPy-NMeth} and \texttt{astropy} \citep{astropy} software packages.\\

NR acknowledges support from UK Science and Technology Facilities Council (STFC) under grant ST/V000594/1. CS acknowledges support from the Agencia Nacional de Investigaci\'on y Desarrollo (ANID) through FONDECYT grant no.\ 11191125 and BASAL project FB210003.
MB is supported by the Polish National Science Center through grants no. 2020/38/E/ST9/00395, 2018/30/E/ST9/00698, 2018/31/G/ST9/03388 and 2020/39/B/ST9/03494, and by the Polish Ministry of Science and Higher Education through grant DIR/WK/2018/12. BG acknowledges the support of the Royal Society through an Enhancement Award (RGF/EA/181006) and the Royal Society of Edinburgh for support through the Saltire Early Career Fellowship (ref. number 1914). CH acknowledges support from the European Research Council under grant number 647112, from the Max Planck Society and the Alexander von Humboldt Foundation in the framework of the Max Planck-Humboldt Research Award endowed by the Federal Ministry of Education and Research, and the UK Science and Technology Facilities Council (STFC) under grant ST/V000594/1.
HH is supported by a Heisenberg grant of the Deutsche Forschungsgemeinschaft (Hi 1495/5-1) as well as an ERC Consolidator Grant (No. 770935). HH acknowledges support from Vici grant 639.043.512, financed by the Netherlands Organisation for Scientific Research (NWO). MH acknowledges support from the National Research Foundation of South Africa (grant no. 137975). JPH gratefully acknowledges support from the estate of George A. and Margaret M. Downsbrough. HYS acknowledges the support from CMS-CSST-2021-A01 and CMS-CSST-2021-B01, NSFC of China under grant 11973070, and Key Research Program of Frontier Sciences, CAS, Grant No. ZDBS-LY-7013. TT acknowledges funding from the Swiss National Science Foundation under the Ambizione project PZ00P2$\_$193352. AHW is supported by an European Research Council Consolidator Grant (No. 770935).
Support for ACT was through the U.S. National Science Foundation through awards AST-0408698, AST- 0965625, and AST-1440226 for the ACT project, as well as awards PHY-0355328, PHY-0855887 and PHY-1214379. Funding was also provided by Princeton University, the University of Pennsylvania, and a Canada Foundation for Innovation (CFI) award to UBC. ACT operated in the Parque Astron\'{o}mico Atacama in northern Chile under the auspices of the Agencia Nacional de Investigaci\'{o}n y Desarrollo (ANID). The development of multichroic detectors and lenses was supported by NASA grants NNX13AE56G and NNX14AB58G. Detector research at NIST was supported by the NIST Innovations in Measurement Science program.
\\
%%%%%%%%%%%%%%%%%%%%%%%%%%
\footnotesize{\textit{Author contributions: All authors contributed to the development and writing of this paper. The authorship list is given in two groups: the lead authors (NCR \& CS) followed by an alphabetical group of those who made significant contributions to the scientific analysis and/or the ACT or KiDS surveys. }}	

\end{acknowledgements}
\bibliographystyle{aa}
\bibliography{master}

\input{appendix}

%\input{appendix}

\end{document}

%% file: appendix.tex
\begin{appendix} 
%\section{Testing the Measurement}
\section{Mis-centering of Clusters}
\label{app}
%\subsection{Mis-centering of Clusters}
The mis-centering of clusters has been reported as one of the largest biases when estimating the cluster mass from weak gravitational lensing \citep[e.g.][]{Sommer2022, Ford2015}. If the true cluster centre has not been identified then the amplitude of the ESD will be reduced at small radii, corresponding to a lower mass estimate for the cluster. 

We consider two different cluster centre definitions: the peak of the SZ emission and the Brightest Cluster Galaxy (BCG), which has been shown to be the best tracer of the cluster centre~\citep{George2012}. The SZ centre is not expected to trace the centre accurately since it is limited by the size of the beam which is $1.4$ arcmin for ACT; this corresponds to $R=0.09 \, {\rm Mpc}$ for $z_{\rm l}=0.1$ and $R=0.62 \, {\rm Mpc}$ for $z_{\rm l}=0.9$. We also anticipate that most clusters are not relaxed and therefore the gas is not expected to trace the centre of the dark matter halo. 

We find the BCG for each cluster by selecting KiDS galaxies found inside a cylinder, centred on the SZ peak, with a length $|z_{\rm B} - z_{\rm l}| = 2 \times 0.04(1+z_{\rm l})$, where $z_{\rm B}$ is the photometric redshift of a KiDS galaxy, and a diameter $2 \times 1.4$ arcmin. The length of the cylinder is defined to be twice the length of the uncertainty on the photometric redshift of the KiDS galaxies, and the diameter is defined based on the size of the ACT beam. From these galaxies the brightest one is chosen. We then confirm the BCG locations by visual inspection, using public imaging from KiDS, the Dark Energy Camera Legacy Survey and Hyper Suprime-Cam. We note that for some clusters (for example one of the most massive clusters in our sample, Abell 2744/ACT-CL J0014.3$-$3022) that are undergoing merging and therefore far from being in a virialised state, it is difficult assign an appropriate centre.  

\begin{figure}
	\includegraphics[width=\columnwidth]{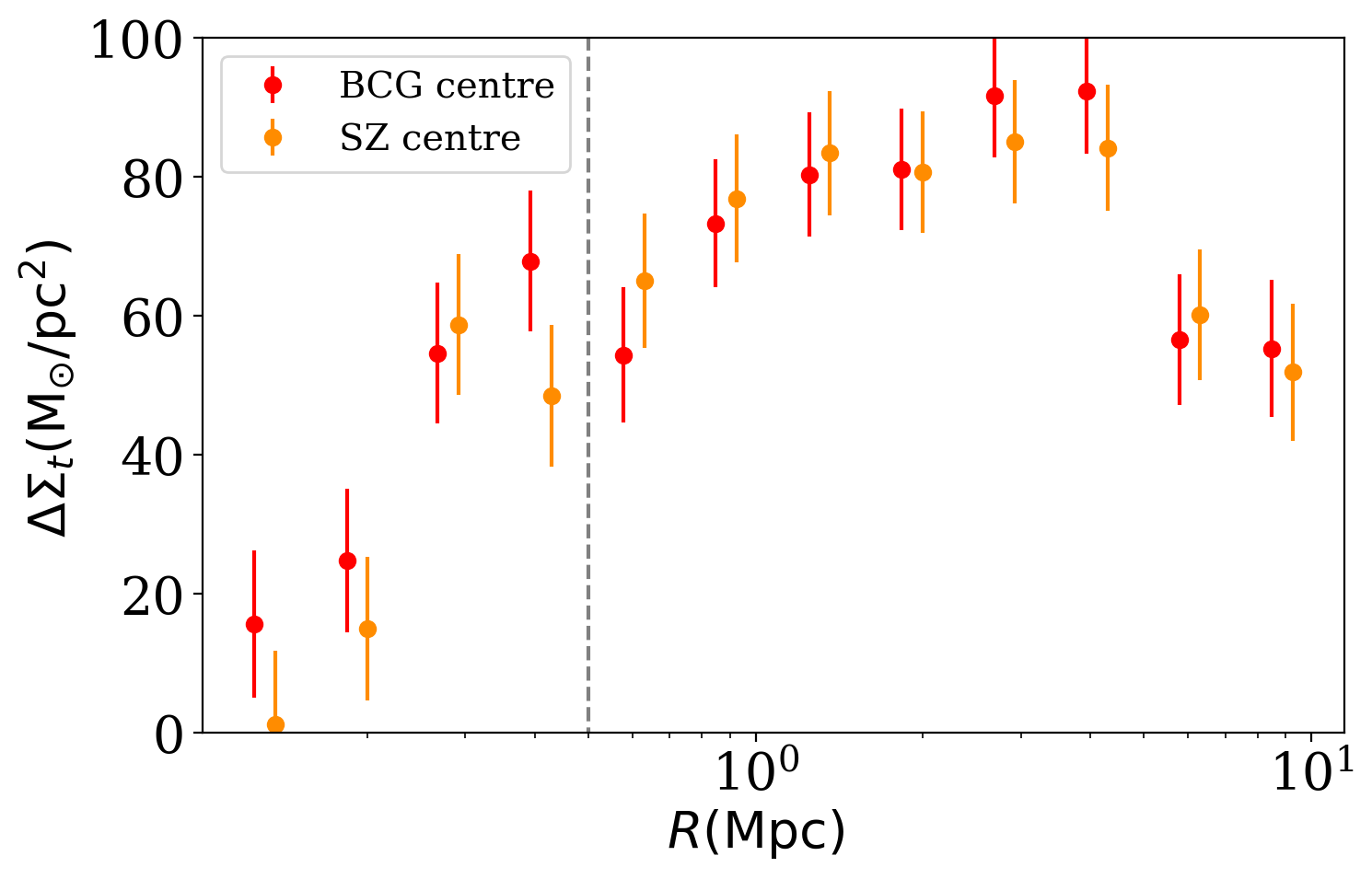}
    \caption{Tests of mis-centering: comparison of the ESD measured around the peak in the SZ emission and the BCG. The ratio between the BCG and SZ centred signal averaged $R>0.5$ is unity.} %\cristobal{\textit{It would be nice to show the best fits for both centres here}}
    \label{fig:miscentering}
\end{figure}

Measuring the ESD around both these centre candidates, we find at small radii the ESD measured around the BCG has a higher amplitude than measured around the SZ centre, which agrees with our expectation -- this is shown in Fig. \ref{fig:miscentering}. 
The ratio between the BCG and SZ centred signal averaged $R>0.5 {\rm Mpc}$ is unity, compared to the signal ratio of $\sim 5$ when averaged $R<0.5 {\rm Mpc}$.  
%I think a referee would let you get away with that. Out of interest when you use the halo model do you recover the same $M$ when using the BCG or SZ centre.  If that is easy to do, that would be a nice thing to add in this appendix as well. }}

%\subsection{Null Tests: B-modes and Random Points}
%To test for systematics in the weak lensing data, we perform two null tests. The cross-component of the ESD, $\Delta \Sigma_{\times}$, is calculated using equation \ref{eqn:ESDest} where the cross--component of shear is defined in equation \ref{eqn:gamma}. This is equivalent to rotating the galaxies by 45 degrees which we expect to remove any tangential alignment. We find $\Delta\Sigma_{\times}$ is consistent with zero; compared to null, this measurement has a p-value of 0.23 and a reduced chi- squared of 1.20. We are therefore confident that no significant systematic effects are present in the galaxy weak lensing data at the scales used in this analysis. 

%We also expect no signal if we measure the ESD at random points across the K-1000 field, unless there is some residual linear bias in the weak lensing data. Compared to null, this measurement has a p-value of 0.42 and a reduced chi- squared of 1.02. Both these measurements are consistent with zero and thus we infer that the signal measured around the ACT cluster positions is due to the over-density of mass.

%\subsection{\naomi{Boost Factor}}
\end{appendix}